\documentstyle[prl,aps,twocolumn]{revtex}

\renewcommand{\(}{\left(}
\renewcommand{\)}{\right)}

\begin{document}
\draft
\title{Creation of Dark Solitons and Vortices in Bose-Einstein Condensates} 

\author
{R. Dum$^{1,2,3}$, J.I. Cirac$^{1,2}$,  M. Lewenstein$^{4}$, and P. Zoller$^{1}$,}
\address{
$^1$Institute f\"ur Theoretische Physik, Universit\"at Innsbruck, \\
A-6020 Innsbruck, Austria}
\address{
$^2$Departamento de F'{\i}sica Aplicada, Universidad de Castilla--La Mancha, 
13071 Ciudad Real, Spain}
\address{
$^3$Ecole Normale Sup\'erieure,
Laboratoire Kastler Brossel,\\
24, Rue Lhomond, F-75231 Paris Cedex 05, France
}
\address{
$^4$ Commissariat \`a l'Energie Atomique, 
DSM/DRECAM/SPAM, Centre d'Etudes de Saclay \\91191 Gif-sur-Yvette, France}

\date{\today}

\maketitle 

\begin{abstract} 
We propose and analyze a scheme to create dark solitons and vortices in
Bose-Einstein condensates. This is achieved starting from a condensate
in the internal state $|a\rangle$ and transferring the atoms to the
internal state $|b\rangle$ via a Raman transition induced by laser
light. By scanning adiabatically the Raman detuning, dark solitons and
vortices are created.
\end{abstract}


Recently Bose-Einstein condensation has been demonstrated in dilute
atomic gases \cite{BEC}. This state of matter resembles other states
found in the fields of superfluidity, superconductivity and nonlinear
optics. It is thus natural to expect that some phenomena that appear in
those fields, such as solitons and vortices, can be observed with dilute
atomic gases. In fact, the Gross--Pitaevskii Equation (GPE) \cite{GP1}
which describes the wavefunction of the macroscopically occupied state
in a trapping potential allows for stationary solutions that represent
dark solitons (see below) and vortices (see also \cite{VORT}). Here we
propose a scheme to generate these solutions in a {\em controlled way}
using an approach based on ``engineering the macroscopic wavefunction'':
once the condensate has formed, we use a coherent Raman process to
create solitons and vortices.

Our idea is to couple the internal state $|a\rangle$ where the
condensate is formed with another internal state $|b\rangle$ using a
Raman transition (Fig.~1). The laser parameters are chosen such that the
state after the transfer is an eigenstate of the GPE
corresponding to solitons or vortices. The complete transfer is achieved
by an adiabatic change of the laser frequency. In the present case,
where the whole process is described by non--linear equations, the
familiar idea of adiabatic transfer along eigenstates has to be taken
with caution since the Hamiltonian describes the atomic interactions by
a mean field depending on the shape of the wavefunctions. We will
analyze two cases: in 1D we will study the creation of dark solitons
and in 2D the creation of vortices.

In the Hartree-Fock approximation the
state $\Phi$ of a condensate of $N$ bosons
confined in a potential $V(\vec r)$ is described by the time-independent GPE
\begin{equation}
\label{GPE}
\left[ -\frac{\hbar^2\vec \nabla^2}{2m} + V(\vec r) 
+ N g |\Phi(\vec r,t)|^2 \right]
\Phi(\vec r,t) = E \Phi(\vec r,t) \; .
\end{equation}
The mean field interactions are characterized by a coupling constant
$g=4\pi\hbar^2a_s/m$, where $a_s>0$ is the $s$--wave scattering length.
In the following analysis we will concentrate on the Thomas--Fermi limit
\cite{BAYM}, since most of the experiments operate in this regime
\cite{BEC}. In this limit, the mean interaction energy is much larger
than the mean kinetic energy which can be neglected when calculating the
ground state solution of the GPE (\ref{GPE}):
\begin{equation} 
\label{Env}
\Phi_{E_0}(\vec r)=\{[E_0-V(\vec r)]/(Ng)\}^{1/2}.
\end{equation}
for $\vec r$ such that $V(\vec r) <E_0$; $E_0$ is determined from
$\int d^3 r |\Phi_{E_0}(\vec r)|^2=1$ which reflects particle
conservation. Here we are interested in other stationary solutions of
(\ref{GPE}) $\Phi(\vec r)$ with energy larger than $E_0$. Far from the
trap center $\vec r=0$ we can still neglect the kinetic energy which
suggests the ansatz
\begin{equation} 
\label{SLOW}
\Phi(\vec r) = \phi(\vec r) \Phi_E(\vec r)
\end{equation}
that is a a product of an envelope function (\ref{Env}) and a function
$\phi$ with the condition $|\phi(\vec r)|\simeq 1$ far away from the
origin. Now $E$ is determined from $\int d^3 r | \phi(\vec r)
\Phi_E(\vec r)|^2=1$. Substituting (\ref{SLOW}) in the GPE and
neglecting the derivatives of $\Phi_E(\vec r)$ we obtain a nonlinear
equation for $\phi(\vec r)$, which near $\vec r=0$, where we can neglect
the variation of the trapping potential, reads 
\begin{equation}
\left[-\frac{\hbar^2\vec \nabla^2}{2m} + N g |\phi(\vec r)|^2 \right] 
\phi(\vec r) = E \phi(\vec r).
\end{equation}
This equation has the same form as the familiar GPE for the homogeneous
case. Together with the boundary conditions $|\phi(\vec r)|\simeq 1$ for
$|\vec r| \rightarrow\infty$ it gives rise to dark solitons and
vortices \cite{RUSSIAN}. Our goal is to design boson-laser interactions which
will generate these solutions.

We assume that the bosons have two internal levels $|a\rangle$ and
$|b\rangle$ as in Rb \cite{RUBI}. The particles interact with a laser
beam that connects these two levels. The evolution of the mean field
spinor $\vec \Phi =(\Phi_a,\Phi_b)$ obeys the following non--linear
equation
\begin{equation}
\label{GPE2}
i\hbar \frac{d}{dt} \vec \Phi(\vec r,t) =
 ({\cal H}+ {\cal H}_1 \vec \Phi(\vec r,t) 
\end{equation}
where
\begin{equation}
\label{Ham1}
{\cal H} = -\frac{\hbar^2\vec \nabla^2}{2m} + V(\vec r) + 
N g [|\Phi_a(\vec r,t)|^2 + |\Phi_b(\vec r,t)|^2],
\end{equation}
describes the linear evolution plus atom--atom interactions and 
\begin{equation}
\label{Ham2}
{\cal H}_1 = \left(
\begin{array}{cc}
0            & \frac{1}{2} \lambda(\vec r) \\
\frac{1}{2} \lambda(\vec r) & -\delta
\end{array}
\right)
\end{equation}
the interaction with the laser. In (\ref{Ham1}) we have assumed that the
interaction between levels $i$ and $j$ ($i,j=a,b$) can be described by a
pseudopotential $g_{ij}\delta(\vec r)$ and that $g_{ij}=g>0$, which
indeed is a good approximation for Rubidium \cite{RUBI}. For $V(\vec r)$
we choose an anisotropic harmonic potential with frequencies
$\omega_{x,y,z}$ which we assume identical for both internal levels. In
(\ref{Ham2}), $\lambda(\vec r)=\Omega_a(\vec r) \Omega_b(\vec
r)/(4\Delta)$ where $\Delta$ is the detuning from the intermediate level
$|r\rangle$, and $\Omega_{a,b}$ are the Rabi frequencies corresponding
to the couplings between $|a\rangle$ and $|b\rangle$ to $|r\rangle$,
respectively (see Fig.~1). Their specific form depends on the laser
configuration. The Raman two--photon detuning is denoted by $\delta$.
The conservation of the number of particles gives the normalization
condition $ \int d^3 \vec r \left[ |\Phi_a(\vec r)|^2 + |\Phi_b(\vec
r)|^2\right] \equiv \pi_a^2 + \pi_b^2 =1$ with $\pi^2_{a,b}$ the
populations in levels $a,b$, respectively.

As the initial state we take $\Phi_{E_0}$ [see Eq.(\ref{Env})] which 
corresponds to the state of the condensate formed in the 
internal level $|a\rangle$. We will design
$\lambda(\vec r)$ and $\delta$ such that the atoms are transferred to
$|b\rangle$ with a wavefunction which corresponds to dark solitons or
vortices. In absence of interactions ($g=0$) the problem reduces to the
one of a single trapped particle. In that case, one can simply use a
resonant $\delta=0$ laser pulse of a well defined area to carry out the
population transfer \cite{IONS}. In presence of interactions, this
method will not work: as soon as particles are transferred to a
different state and the shape of the wavefunction changes, the
interaction energy changes [Fig.~1(a,b)]. Therefore, an initially
resonant Raman laser pulse at the beginning of the pulse soon becomes
off--resonant, and the transfer process will stop. We circumvent this
problem by using adiabatic passage. The idea is to start from a {\it
negative} Raman detuning so that the atoms do not feel the laser
[Fig.~1(a)]. Then, the Raman detuning is changed adiabatically to
sufficiently large {\it positive} values [Fig.~1(b)]. As soon as the
laser frequency approaches the Raman resonance, the atoms will start
flowing to the state $|b\rangle$. The fact that the interaction energy
changes will effectively change the value of $\delta$ at which this
resonance occurs. This will not affect the overall process provided the
final value of $\delta$ is large enough so that at the end the atoms do
not feel the off--resonant laser anymore. The reason is that adiabatic
transfer only depends on the initial and final values of the adiabatic parameter
(detuning). 

In order to describe analytically the adiabatic process we look for
stationary solutions $\vec \Phi_\delta(\vec r)$ of (\ref{GPE2}) for a
given value of $\delta$. The idea is to change $\delta$ adiabatically so
that the state of the system changes according to $\vec \Phi_\delta$ in
accordance to the adiabatic theorem. We thus have to impose that for the
initial and the final Raman detunings $\delta_0$ and $\delta_f$ the
spinor $\vec \Phi_\delta$ corresponds to the initial state and the
desired state, respectively. That is: (i) for $\delta \rightarrow
\delta_0$, $\vec \Phi_\delta \simeq\(\Phi_{E_0},0 \)$; (ii) for
$\delta\rightarrow \delta_f$, $\vec \Phi_\delta =\(0,\Phi_b\)$ where
$\Phi_b$ is another stationary solution of Eq.~(\ref{GPE}). In order to
find the appropriate stationary solutions of (\ref{GPE2}) we have to
solve first the set of nonlinear equations
\begin{equation}
\label{HPhi}
{\cal H} \ \Phi^{\pi_a}_{a,b}(\vec r) = \epsilon_{a,b}^{\pi_a}
\Phi^{\pi_a}_{a,b}(\vec r),
\end{equation}
for a fixed (real) value of $\pi_a$ (with $\pi_b=|1-\pi_a^2|^{1/2}$ due
to normalization) and satisfying the conditions (i,ii). Here
$\Phi^{\pi_a}_{a,b}$ describe the wavefunctions in absence of laser
coupling for given populations $\pi_{a,b}$. The presence of the laser
will lead to a dressing of these levels in analogy with the well-known
picture known from non-interacting atoms in presence of laser light
\cite{CLAUDE}. Once these functions are found, we replace ${\cal H} \vec
\Phi$ in (\ref{GPE2}) by (\ref{HPhi}), that is we restrict the evolution
to the subspace defined by these two functions; by multiplying the first
equation by $\Phi_a(\vec r)^\ast$ and the second by $\Phi_b(\vec
r)^\ast$ and integrating we obtain
\begin{equation}
\label{Dressed}
\left(\begin{array}{cc} \epsilon_a^{\pi_a} & \lambda^{\pi_a} \\ 
   \lambda^{\pi_a \ast} & \epsilon_b^{\pi_a} - \delta \end{array}\right) 
\left(\begin{array}{c} \pi_a \\ \pi_b \end{array}\right) = 
E
\left(\begin{array}{c} \pi_a \\ \pi_b \end{array}\right),
\end{equation}
where we have defined 
\begin{equation}
\label{lambda}
\lambda^{\pi_a}= \frac{1}{2\pi_a\pi_b} \int d^3\vec r\, \lambda(\vec r)\,
\Phi^{\pi_a}_a(\vec r)^\ast \,\Phi^{\pi_a}_b(\vec r).
\end{equation}
Equations (\ref{Dressed}) define two generalized dressed states of our
system which take fully into account atomic interactions. From these
equations we can determine the values of $\delta$ and $E$ corresponding
to $\Phi^{\pi_a}_{a,b}$. In summary, the problem is reduced to solve the
coupled eigenvalue equations (\ref{HPhi}) for a given value of $\pi_a$,
such that varying continuously this parameter we go from $\Phi_a(\vec
r)=\Phi_{E_0}(\vec r)$ to the desired state $\Phi_b(\vec r)$. The energy
separation $\Delta_{a,b}(\pi_a) \equiv \epsilon_b^{\pi_a} -
\epsilon_a^{\pi_a}$ between the bare wavefunctions $\Phi^{\pi_a}_{a,b}$
gives the resonance condition for the Raman detuning, $\delta\rightarrow
\Delta_{a,b}(\pi_a)$, for fixed level populations $\pi_{a,b}$. The
induced width is given by $\lambda(\pi_a)$ for $\pi_a^2\simeq 1/2$ which
gives the avoided crossing and therefore the time scale for
adiabaticity. 

We illustrate this procedure now for the 1D case. This corresponds to
the limit in which $\omega_{x,y} \gg \omega_z$ so that the dynamics
along the $x$ and $y$ direction is frozen. Our goal is to create a dark
soliton starting from $\Phi_{E_0}(z)$. We are interested in dark
solitons with a zero at the trap center. This requires that the laser
interaction changes the parity of the wavefunction when the atoms are
transferred from a to b. To this aim we choose the simplest laser
configuration, so that $\lambda(z)=\lambda_0 \sin (kz)$, i.e. a standing
wave. In order to achieve an efficient coupling we take $k \le 1/z_0$,
where $z_0$ is the size of the $\Phi_{E_0}$ \cite{EVANESCENT}; note that
the effective $\lambda$ defined in (\ref{lambda}) will be very small if
$k z_0 \gg 1$. In this case the avoided crossing of the dressed energy
levels will be of the order $\lambda_0$, which sets the time scale for
the adiabaticity. On the other hand, $\lambda_0$ has to be smaller than
the typical energy separations $|\Delta_{ab}|$ so that the Stark shifts
do not mix these wavefunctions with others of higher energies
$\epsilon$. The initial value of the Raman detuning must be $\delta_0
\ll \Delta_{a,b}(\pi_a=1)$, whereas the final value must fulfill
$\delta_f \gg \Delta_{a,b}(\pi_a=0)$. In Fig.~2 we have plotted
numerical results of the solutions of the time--dependent GPE
(\ref{GPE2}). Figure 2(a) shows the spatial distribution
$P_{a,b}(z)=|\Phi_{a,b}(z)|^2$ corresponding to states $|a\rangle$
(solid line) and $|b\rangle$ (dashed line). As the transfer progresses,
we see that the wavefunction of the atoms in $|a\rangle$ narrows and the
one of the atoms in $|b\rangle$ develops a hole in the center. When most
of the atoms are in b, the corresponding wavefunction contains a dark
soliton. This manifests itself in the effective (trap plus mean field)
potential; it is initially flat, and later it develops a narrow dip as a
consequence of the dark soliton [see Fig.~1(b)]. The part of the atoms
still in $|a\rangle$ become trapped in a bound state of this dip, which
becomes deeper as we move more atoms in the excited state. In Fig.~2(b)
we have plotted the fraction of atoms $P_{a,b}=\pi_{a,b}^2$ of levels a
and b. As this figure shows, the transfer efficiency is essentially
$100$ percent. For the analytical understanding of these results, we
proceed as explained above in terms of generalized dressed states.
First, in the limit $\pi_a\rightarrow 1$ one can estimate the value of
$\Delta_{ab}$ using a square well of length equal to the size of the
Thomas Fermi solution $\Phi_{E_0}$. In the opposite limit,
$\pi_a\rightarrow 0$ we can calculate $\Delta_{ab}$ in the same way as
in the context of Eq.~(\ref{Env}). We write $\vec \Phi(z)
=[\phi_a(z),\phi_b(z)] \Phi_E(z)$, with $\Phi_E(z)$ defined in
(\ref{Env}). Near the trap center we obtain 
\begin{equation} 
\label{1D}
\left[-\frac{\hbar^2}{2m}\frac{d^2\ }{dz^2} + N g |\phi_a(z)|^2
+\epsilon_b |\phi_b(z)|^2   \right] \phi_{a,b}(z) = \epsilon_{a,b}
\phi_{a,b}(z), 
\end{equation}
with the boundary conditions $|\phi_a(z)|\rightarrow 0$ and
$|\phi_b(z)|\rightarrow 1$ as $|z|\rightarrow \infty$. We find that
$\phi_a(z)=A {\rm sech} (\alpha z)$ and $\phi_b(z)=\tanh (\alpha z)$
solve Eq.~(\ref{1D}). With the normalization condition we can find the
values of $A$, $\alpha$ and $\epsilon_{a,b}$ for a given value of
$\pi_a$. This solution is in perfect agreement with our 1D numerical
results. We performed a full $3D$ integration of the GPE in order to
make sure that the $1D$ effect is not effected by the presence of the
transverse degrees of freedom \cite{1D3D}.

A two-soliton solution is obtained by starting from the ground state in
a, and coupling with a laser configuration which preserves the parity,
$\lambda(z)=\lambda_0 \cos (kz)$. In order not to couple to the ground
state in b, the initial detuning has to be $\delta_0 > |\lambda_0|$. We
then increase the detuning adiabatically to a sufficiently large value.
In Fig. ~3 we show plots of numerical solutions. At the end of the
process all the particles are in the state $|b\rangle$ with a
wavefunction that includes two dark solitons. Again an analytical Ansatz
is possible: for $|\pi_b|\rightarrow 1$ we set
$\Phi_b(z)=\phi_b(z)\Phi_E(z)$ with $\phi_b(z)= \tanh[\alpha(z-a)]
\tanh[\alpha(z+a)]$ ($a$ is a free parameter) which reproduces the
numerical results very well.
 
A 2D situation arises in the limit $\omega_z \gg
\omega_\perp=\omega_x=\omega_y$. In this limit we are interested in
creating vortex solutions of the form $\Phi(\rho,\varphi)= f(\rho)
e^{i\varphi}$ where $\rho$ and $\varphi$ are cylindrical coordinates and
$f(\rho)$ is a function with a zero at $\rho=0$. In order to provide the
required angular momentum to the atoms that are transferred, we choose a
laser configuration such that $\lambda(x,y)=\lambda_0 [\sin(k_L x) + i
\sin(k_L y)] \simeq \lambda_0 k_L\rho e^{i\varphi}$ for $k_L\rho \alt 1$
\cite{ANGULAR}. The density distribution $|\Phi(x,y)|^2$ after an
adiabatic switch of the detuning is plotted in Fig.~4. The insert shows
that all the population is transferred to the vortex state. Analytical
approximations can be obtained with the ansatz $\phi_{a}(\vec r)= A \,
{\rm sech}(\alpha\rho)$ and $\phi_b(\vec r)= \tanh(\alpha\rho)
e^{i\varphi}$. 

A simple way of observing the shape of the density $n(\vec r)$ is by
opening of the trap. In a way similar to Ref.~\cite{YVAN} we can show
that the density at later times is related to the density at $t=0$ by
$n(\vec r,t)=n[\vec r/\gamma(t),0]/\gamma^{2 d}(t)$ where 
the scaling factors obey $\ddot{\lambda}=\omega^2/\lambda^{d+1}$ ($d$ is
the dimension) and initial conditions $\lambda=1,\dot{\lambda}=0$. This
leads to an asymptotic behavior $\gamma(t) \rightarrow \sqrt{2}
\omega_z t$ for the 1D dark soliton and $\gamma(t) \rightarrow
\omega_\perp t$ for the vortex solution. This selfsimilar expansion
without a change in shape is typical for solitonic behavior. On the
other hand, another important issue to address is the stability of
vortices and dark solitons. In some recent works dealing with
excitations of vortex states \cite{VORT} it is shown that quasiparticles
states localized near the center of the trap will be preferentially
occupied by collisions thereby destabilizing the vortex. However, using
the analogy between our proposal and the excitations of condensates via
time--dependent trapping potentials \cite{CD}, we expect the destabilization
time to be much longer than that required for the creation of vortices
and solitons. The reason is that the creation of vortices and dark
solitons involve regular (non--chaotic) solutions of the GPE, which
give rise to destabilizations that grow only polynomially with time
(instead of exponentially) \cite{CD}.

We have demonstrated that one can engineer the macroscopic wavefunction
of Bose--Einstein condensed sample by coupling the internal atomic
levels with a laser. The method is based on  adiabatic transfer of
population along generalized dressed states which include the nonlinear
atom--atom interactions. We find the technique to be very robust against
uncertainties in the parameters of the problem. Furthermore, we have developed
analytical approximations to describe this process, and show with
explicit examples how to generate dark solitons and vortices. Our
numerical results confirm our predictions and demonstrate the stability
of these solutions. We expect that this method can be applied to current
or planned experiments.

We thank K. Burnett for discussions. This work was supported in part by
the TMR network ERB--FMRX--CT96--0087, and by the Austrian Science
Foundation.

\begin{figure} 
%\centerline{\epsfysize=65mm\epsfbox{fig1.eps}}
\vspace{5mm}
\caption{ 
Schematic representation of the process: (a) initial state; (b) final state.
}
\end{figure}

\begin{figure}
%\centerline{\epsfysize=65mm\epsfbox{fig2.eps}}
\vspace{5mm}
\caption{Generation of a dark soliton. The detuning is varied linearly
with time from $\delta=-1.5\omega$ to $\delta=6.5\omega$. Other parameters:
$\lambda_0=0.15\omega$, $k_L=0.5/a_0$ and 
$Ng=50 \hbar\omega a_0$, where $a_0^2=\hbar/(m\omega)$. (a) snap shots
of the position distributions of the wavefunctions corresponding to
atoms in level $|a\rangle$ (solid line) and $|b\rangle$ (dashed
line) for different times; (b) Populations of these levels as a function of time. 
}
\end{figure}

\begin{figure}
%\centerline{\epsfysize=65mm\epsfbox{fig2.eps}}
\vspace{5mm}
\caption{Generation of two dark solitons. The detuning is varied linearly
with time from $\delta=0.25\omega$ to $\delta=5\omega$. Other parameters:
$\lambda_0=0.15\omega$, $k_L=0.5/a_0$ and 
$Ng=20 \hbar\omega a_0$, where $a_0^2=\hbar/(m\omega)$. 
}
\end{figure}

\begin{figure}
%\centerline{\epsfysize=65mm\epsfbox{fig3.eps}}
\vspace{5mm}
\caption{\label{2D}
Generation of a vortex: position distribution of the final state.
The detuning is varied linearly from 
$\delta=-0.6\omega_\perp$ to $\delta=5\omega_\perp$.
The parameters are $Ng=500 \hbar \omega_\perp a_\perp^2$,
$\lambda=0.15\hbar\omega_\perp$ and $k_L= 0.5 a_\perp$. The inset
shows the evolution of the populations in levels $|a\rangle$ and 
$|b\rangle$ (solid and dashed lines, respectively).
}
\end{figure}


\begin{thebibliography}{99}

\bibitem{BEC}  
M. H. Anderson et al.,
Science {\bf 269}, 198 (1995);
K. Davis et al., Phys. Rev. Lett. {\bf 75}, 3969 (1995);
C.C. Bradley et al., Phys. Rev. Lett. {\bf 78}, 985 (1997).

\bibitem{GP1}
M. Edwards et al., Phys. Rev. A {\bf 53}, R1950 (1996);
F. Dalfovo, S. Stringari, Phys. Rev A {\bf 53}, 2477 (1996).

\bibitem{VORT} R.J. Dodd et al., Phys. Rev A {\bf 56}, 587 (1997);
D.S. Rokhsar, Phys. Rev. Lett. {\bf 79}, 2164 (1997).

\bibitem{BAYM} G. Baym and C. Pethick, Phys. Rev. Lett.  {\bf 76}, 6 (1996).

\bibitem{RUSSIAN} V.E. Zakharov, A.B. Shabat, JETP {\bf 37}, 823 (1973);

\bibitem{RUBI}  C.J. Myatt et al.,Phys. Rev. Lett. {\bf 78}, 586 (1997).

\bibitem{IONS}
See, e.g., 
J. I. Cirac {\it et al.}, Adv. At. Mol. Phys. {\bf 37}, 237 (1996).

\bibitem{CLAUDE} C. Cohen-Tannoudji et al., Photon-Atom Interactions
(Wiley 1992).

\bibitem{EVANESCENT} 
For typical experimental conditions 
a typical wavelength is $500 nm$ and $z_0$ is
$\approx 5 \mu m$; therefore one has to use two laser beams forming
a small angle to achieve $k_Lz_0\simeq 1$.

\bibitem{1D3D}
We take $\omega_z/\omega_\perp=0.1$ and we relate the interactions by
$g_{1D}={1 \over \pi} {g_{3D} \over a_\perp^2}$ where
$a_\perp=(\hbar/m\omega_\perp)^{1/2}$.

\bibitem{ANGULAR} 
A coupling of this form can be achieved using Raman beams with
wave vectors $\vec{k}_{1,2}$ such that $|\vec{k}_1-\vec{k}_2| <k_L$.
The cross terms with wavenumber $\vec{k_1}+\vec{k_2}$ can be neglected due
to Lamb-Dicke suppression.

\bibitem{YVAN} 
Y. Castin and R. Dum, Phys. Rev. Lett.  {\bf 77}, 5315 (1996).

\bibitem{CD}
Y. Castin and R. Dum, Phys. Rev. Lett. {\bf 79}, XXXX (1997).

\end{thebibliography}
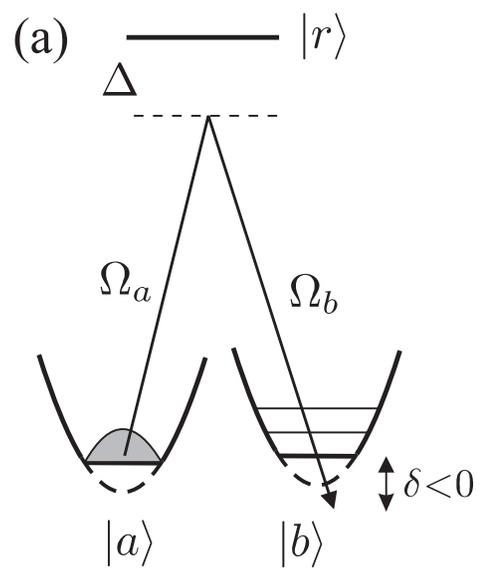
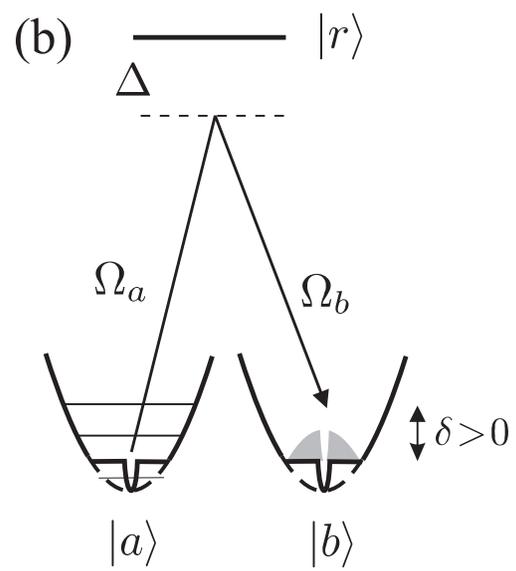
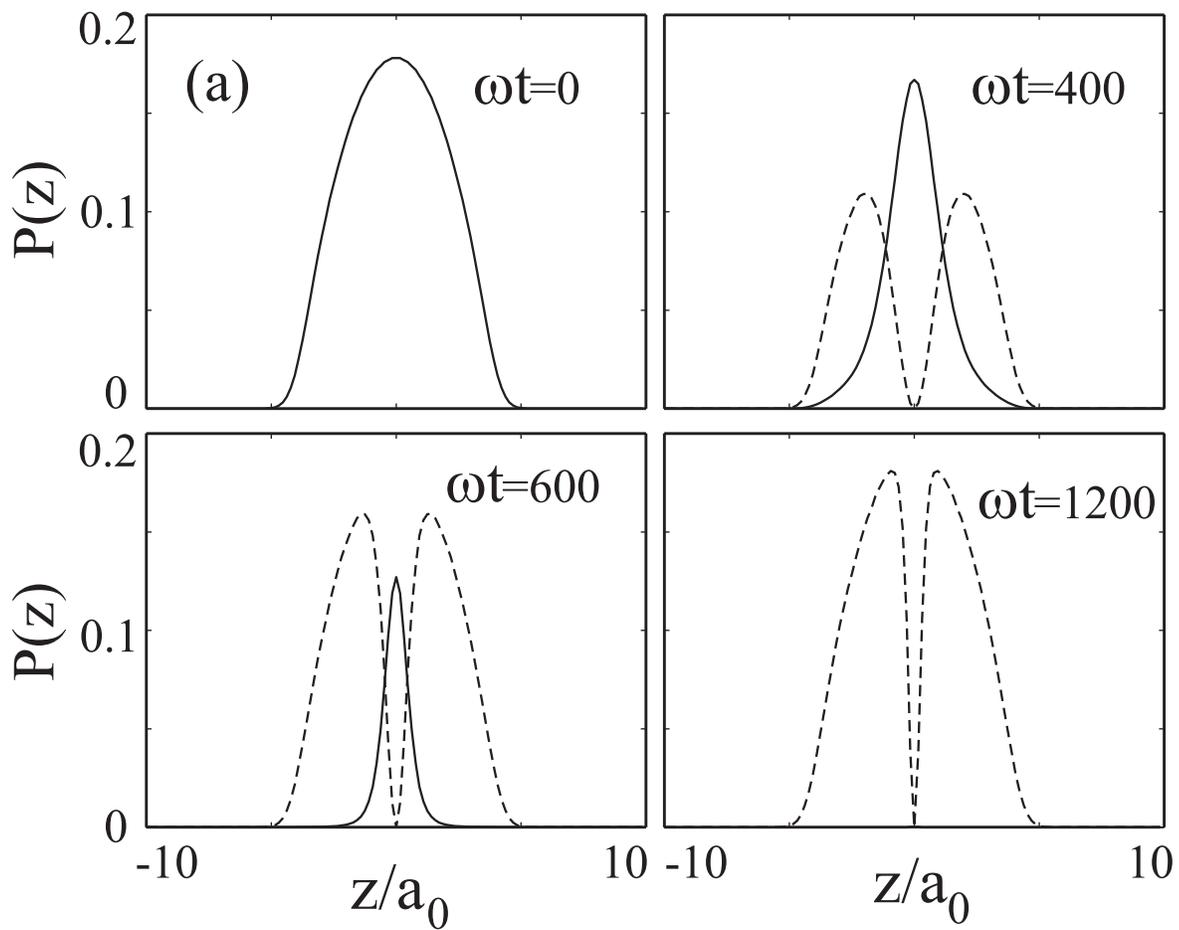
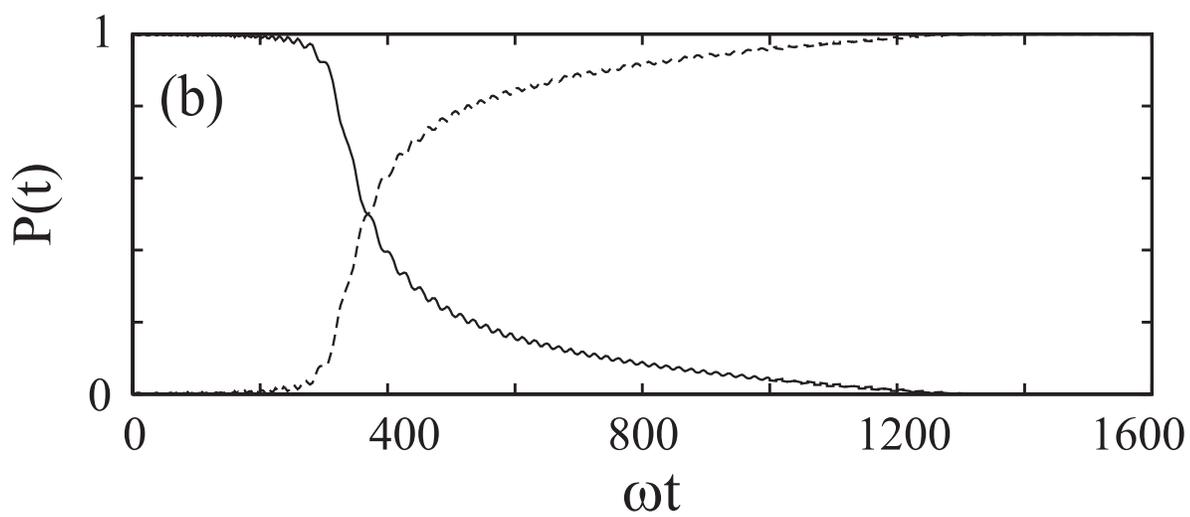
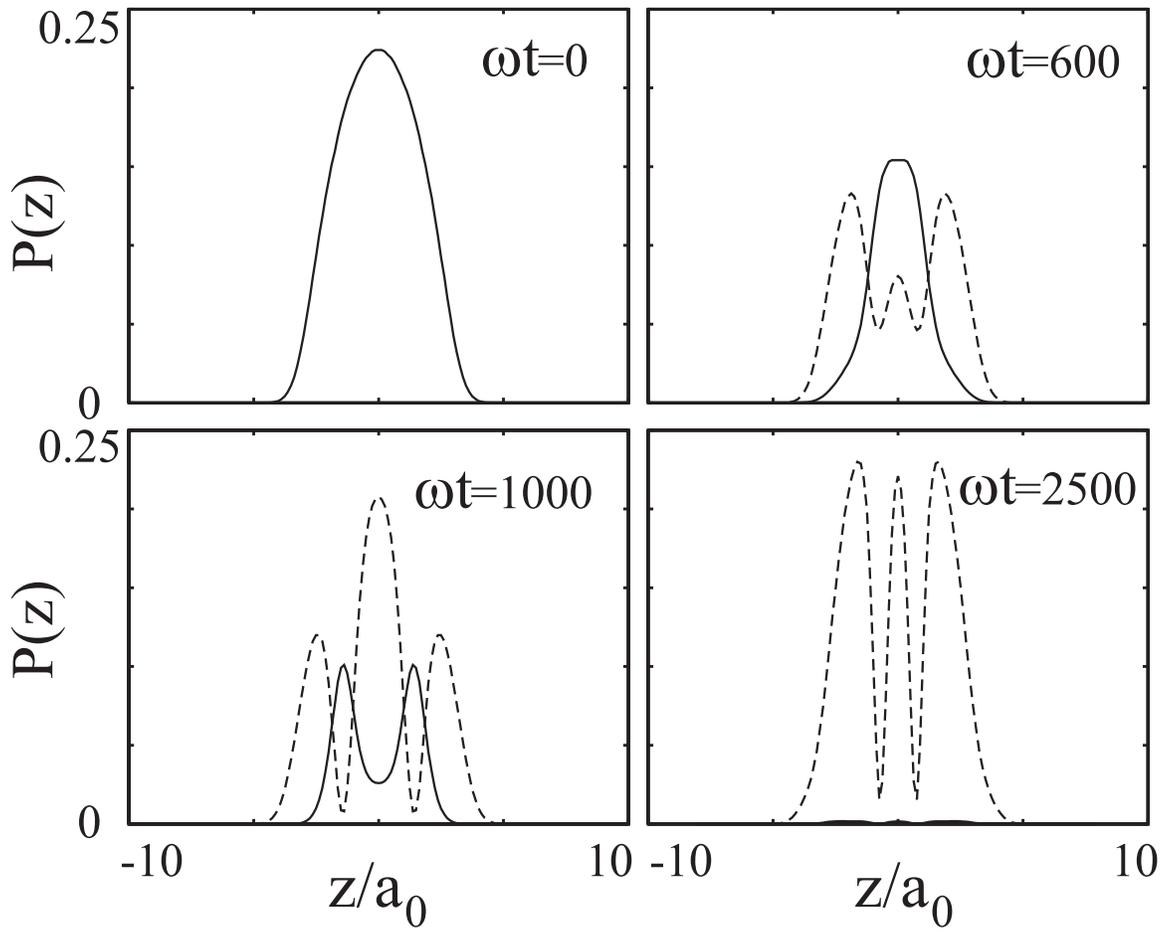
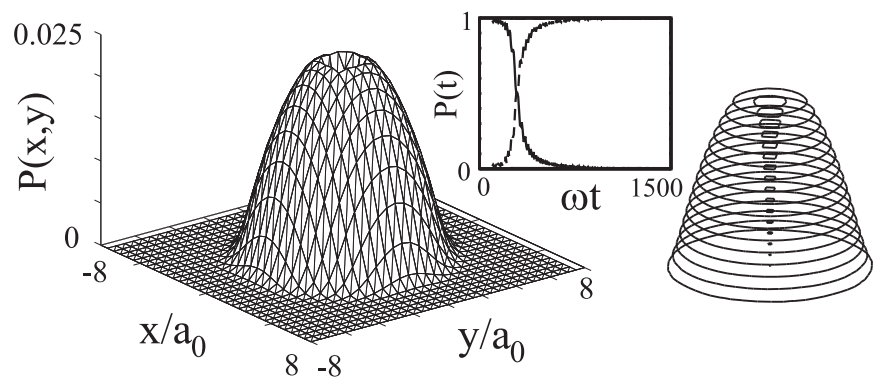
\end{document}